\documentclass{PoS}

\usepackage{cancel}

\usepackage{subfig}

\usepackage{amsmath}

\usepackage{amssymb}
\usepackage{latexsym}
\usepackage{wasysym}
\usepackage{pstricks}

\newlength{\myl}
\newcommand{\msbar}{\overline{\mbox{MS}}}

\newcommand{\bga}{\begin{gather}}

\newcommand{\ega}{\end{gather}}

\newcommand{\bal}{\begin{align}}

\newcommand{\eal}{\end{align}}

\newcommand{\z}{\zeta}

\newcommand{\ice}[1]{{}}

\newcommand{\EQN}[1]{\label{#1}}

\newcommand{\ed}{\end{document}}

\newcommand{\prd}{\partial}
\newcommand{\ep}{\epsilon}
\newcommand{\beq}{\begin{equation}}
\newcommand{\eeq}{\end{equation}}
\newcommand{\bea}{\begin{eqnarray}}
\newcommand{\eea}{\end{eqnarray}}

\newcommand{\ba}{\begin{array}}
\newcommand{\ea}{\end{array}}

\newcommand{\g}{\gamma}

\newcommand{\be}{\beta}
\newcommand{\bc}{\begin{center}}
\newcommand{\ec}{\end{center}}

\newcommand{\re}[1]{(\ref{#1})}

\newcommand{\unb}[1]{\underbrace{#1}}

\catcode`\@=11
\def\slash{\mathpalette\make@slash}
\def\make@slash#1#2{\setbox\z@\hbox{$#1#2$}%
  \hbox to 0pt{\hss$#1/$\hss\kern-\wd0}\box0}
\catcode`\@=12 

\def\bbuildrel#1_#2^#3%
{\mathrel{\mathop{\kern 0pt#1}\limits_{#2}^{#3}}}

\newcommand{\nnb}{\nonumber}

\def\beq{\begin{equation}}
\def\eeq{\end{equation}}
\def\bea{\begin{eqnarray}}
\def\eea{\end{eqnarray}}
\def\bq{\begin{quote}}
\def\eq{\end{quote}}

\def\nnb{\nonumber}

\def\nnb{\nonumber}

\def\ba{\begin{array}}
\def\ea{\end{array}}

\def\bbuildrel#1_#2^#3%
{\mathrel{\mathop{\kern 0pt#1}\limits_{#2}^{#3}}}

\newcommand{\lQ}{\ell_\mu}

\newcommand{\bfl}{\begin{flalign}}

\newcommand{\defas}{\mathrel{\mathop:}=}
\newcommand{\mzv}[2][]{\zeta^{#1}_{#2 }}

\newcommand{\hz}{\hat{\zeta}}

\newcommand{\mbib}{ 

\bibliographystyle{JHEP}
\bibliography{beta5,lit,chet,gm5,rg,dim_reg,JJ,%
steinhauser,baikov,asmirnov,smirnov,vladimirov,vermaseren,%
surguladze,laporta,gorishnii,tarasov,bierenbaum,kataev,czakon,kazakov,%
broadhurst,gracey,%
DIS,DIS2,DIS4,acat,maszeta_PB}
\ed
}

\title{No-$\pi$ Theorem for Euclidean Massless Correlators}

\ShortTitle{No-$\pi$ Theorem }

\author{P.~A.~Baikov,\\
       Skobeltsyn Institute of Nuclear Physics, Lomonosov Moscow State University, 
      1(2), \\Leninskie gory, Moscow  119991, Russian Federation
      \\
        E-mail: \email{baikov@theory.sinp.msu.ru}}

\author{\speaker{K.~G.~  Chetyrkin}%
         \ice{\thanks{A footnote may follow.}}
\\
  II. Institut f\"ur Theoretische Physik
  Universit\"at Hamburg
   Luruper Chaussee 149,\\ 
  22761 Hamburg,  Germany
\\
E-mail : \email{konstantin.chetyrkin@desy.de} }

\abstract{
We provide the  reader with a (very)  short review of recent advances 
in our understanding of the $\pi$-dependent terms in massless
(Euclidean) 2-point functions  as well as in  generic 
anomalous dimensions and $\beta$-functions. We extend the considerations of 
\cite{Baikov:2018wgs} by one more loop, that is   for the case of 6-loop  correlators 
and  7-loop renormalization group (RG) functions.    
}

\FullConference{Loops and Legs in Quantum Field Theory (LL2018)\\
		29 April 2018 - 04 May 2018\\
		St. Goar, Germany}

\begin{document}

\section{Introduction and Preliminaries}


Since the seminal calculation of the Adler function at order $\alpha_s^3$
\cite{Gorishnii:1991vf} it has been known that p-functions
demonstrate  striking   regularities in \ice{as for} 
terms proportional to $\pi^{2 n}$,  with  $n$ being positive  integer.
Here by  p-functions we understand ($\msbar$-renormalized) Euclidean 
Green  functions\footnote{Like quark-quark-qluon vertex in QCD with
  the external gluon line carrying no momentum.}
or 2-point correlators or even some combination thereof,
expressible in terms of massless propagator-like Feynman integrals
(to be named p-integrals below).

To describe  these regularities   we need  to
introduce a few notations and conventions. (In what  follows we limit ourselves
by the case of QCD considered in the Landau gauge).
Let 
\beq
F_n(a,\lQ)  = 1  +\sum_{ 1\le i \le  n }^{0 \le j \le i} g_{i,j} \, (\ell_{\mu})^j \, a^i
\EQN{G:def}
{}
\eeq
be a p-function, 
where $a=\frac{\alpha_s(\mu)}{4\,\pi}$, $\lQ= \ln \frac{\mu^2}{Q^2}$ and $Q$ is
an (Euclidean) external momentum. The integer $n$ stands for the (maximal)
power  of $\alpha_s$ appearing in the p-integrals  contributing to
$F_n$. 
The $F$ without $n$ will stand as a shortcut for a formal series $F_\infty$.
In terms of bare quantities\footnote{We assume  the use of  the dimensional regularization  with the space-time dimension
$D=4-2\,\ep$.} 
\beq
F = Z\, F_B(a_B,\lQ), \hspace{2cm} Z  = 1 +\sum_{ i \ge 1}^{1 \le j \le i} 
Z_{i,j} \, \frac{a^i}{\ep^j}
\EQN{G:GB}
{},
\eeq
with   the bare coupling  constant and the corresponding renormalization constant being
\beq
a_B = \mu^{2\ep} Z_a \,a , \hspace{2cm}  Z_a   = 1 +\sum_{  i \ge 1 }^{1 \le j \le i} 
\Bigl(Z_a\Bigr)_{i,j} \, \frac{a^i}{\ep^j}
{},
\eeq
\ice{
Here $\ep$ is the standard parameter of dimensional regularization 
related with the running space-time dimension $D$ via $D=4-2\,\ep$.
The evolution equation for F reads:
}
\beq
\Bigl(\frac{\prd}{\prd \lQ}\, + \be\,a\, \frac{\prd}{\prd a}\Bigr) F = \g\, F
\EQN{G:RG:evol}
{},
\eeq
with the anomalous dimension (AD) 
\beq
\g(a) =  \sum_{ i \ge 1 } \g_i \, a^i, \ \    \g_i = -i Z_{i,1}  
\EQN{gama:def}
{}.
\eeq
The coefficients of the $\beta$-function $\beta_i$ are related to  $Z_a$ in the 
standard way:
\beq
\beta_i = i \left(Z_a\right)_{i,1} 
\EQN{beta:coef:def}
{}.
\eeq
A p-function $F$ is called scale-independent if the corresponding AD $\gamma \equiv 0$.  
If $\gamma \not=0$ then one  can always construct a scale-invariant object from
 $F$ and $\g$, namely\ice{ 
The  trick was first applied for constructing a $\pi$-free version of a
 2-point  correlator in \cite{Vermaseren:1997fq}.
The same quantity as defined in \re{dG1} was discussed in  \cite{Kubo:1983ws} under the name of the
``Renormalization Scheme invariant anomalous dimension''. 
In DIS similar objects are called "physical anomalous
dimensions", see, e.g. \cite{Davies:2017hyl}. 
}:
\beq
F^{\mathrm si}_{n+1}(a,\lQ) = \frac{ \prd }{\prd \lQ} \left(\ln F\right)_{n+1}
\equiv \Biggl(\frac{\left( \g(a) - \beta(a) a \frac{\prd}{\prd a}\right)\,F_n}{F_n}\Biggr)_{n+1}
{}.
\EQN{dG1}
\eeq         
Note that $F^{\mathrm si}_{n+1}(a,\lQ)$ starts from the first power of the coupling
constant $a$ and is formally   composed from ${\cal O}(\alpha_s^{n+1})$ Feynman  diagrams.
In the same time  is can be completely restored from $F_n$ and the  $(n+1)$-loop AD $\g$.

An (incomplete) list of the  currently known  regularities\footnote{
For discussion of particular examples of  $\pi$-dependent contributions 
into various p-functions we refer  to works \cite{Jamin:2017mul,Davies:2017hyl,Chetyrkin:2017bjc,Ruijl:2018poj}.}
 includes the  following cases.
\ice{
(for
gauge-dependent p-functions we assume  the use of the  Landau gauge). 
}
\begin{enumerate}
\item  Scale-independent p-functions $F_n$ and $F^{\mathrm si}_{n}$ with $n \le 4$  are  free 
from $\pi$-dependent terms.
\item  Scale-independent p-functions $F^{\mathrm si}_5$ are free from $\pi^6$ and $\pi^2$ 
but do depend on $\pi^4$.
\item The QCD $\beta$-function starts  to depend on $\pi$ at  5 loops  only 
\cite{Baikov:2016tgj,Herzog:2017ohr,Luthe:2017ttg}
(via $\z_4 = \pi^4/90$).
In addition, there  exits a remarkable identity \cite{Baikov:2018wgs}
\[
\beta_5^{\z_4} = \frac{9}{8}  \beta_1\, \beta_4^{\z_3}, \    \  \ 
\mbox{with} \  \  \ F ^{\z_i} = \lim_{\z_i \to 0} \frac{\prd}{\prd \z_i} F 
{}.
\]
\ice{
where 
the upper-script $\z_i$  means
$
 F ^{\z_i} = \lim_{\z_i \to 0} \frac{\prd}{\prd \z_i} F 
\ \ \mbox{and (for future reference)} \ \  
F ^{\z_i \z_j} = \lim_{\z_i \to 0} \frac{\prd}{\prd \z_i} F^{\z_j} 
{}.
$
}
\item  If we change   the $\msbar$-renormalization scheme as follows:
\beq
a = \bar{a} \, ( 1 + c_1\, \bar{a} +c_2\,  \bar{a}^2 + c_3 \, \bar{a}^3 
+ \frac{1}{3} \, \frac{\beta_5}{\beta_1}\, \bar{a}^4 ) 
\EQN{new:scheme:1}
{},
\eeq
with $c_1, c_2$ and  $c_3$ being any rational numbers, 
then the function $\hat{F}^{\, \mathrm si}_5(\bar{a},\lQ)$ and  the (5-loop) $\beta$-function
$\bar{\beta}(\bar{a})$ both loose any dependence on  $\pi$. \mbox{This remarkable fact  was discovered 
in \cite{Jamin:2017mul}.}
\end{enumerate}
It should be stressed that  eventually  every separate diagram contributing to 
$F_n$ and $F_{n+1}$ contains the following set of irrational numbers:
$\z_3, \z_4, \z_5, \z_6$ and $\z_7$ for $n =4$,   $\z_3$, $\z_4$ and $\z_5$ for $n=3$.
Thus,   the regularities listed above are quite nontrivial and  for sure  can not be explained by pure
coincidence.  

\section{Hatted representation   of p-integrals and its implications}

The full understanding and  a generic proof of points 1,2 and 3 above have been recently achieved in our work
\cite{Baikov:2018wgs}. 
The main tool of the work was  the so-called ``hatted'' 
representation  of  transcendental objects contributing to a 
given set of p-integrals. Let us  reformulate   the main results of \cite{Baikov:2018wgs} in an abstract form.


We will call  the set  of all  L-loop p-integrals ${\cal P}_L$  a $\pi$-safe one if   the following is  true.

(i)  All p-integrals from the set can  be expressed in terms of $(M + 1)$ 
mutually independent  (and $\ep$-independent) transcendental generators 
\beq
{\cal T}=\{t_1,t_2, \dots , t_{M+1}\}\ \ \mbox{with} \ t_{M+1} = \pi
{}.
\eeq
This means that any p-integral $F(\ep)$ from    ${\cal P}_L$ can be {uniquely}\footnote{
We assume that $F(\ep, {t}_1,{t}_2, \dots , \pi)$ does not  contain  terms  proportional to $\ep^n$
with $n \ge 1$.}
 presented  as follows 
\beq F(\ep) = F(\ep, {t}_1,{t}_2, \dots , \pi) + {\cal{ O}}(\ep)
\label{Fep}
{},
\eeq
\ice{
\beq
F(\ep) = F(\ep, \hat{t}_1,\hat{t}_2, \dots ,\hat{t}_M, \pi) + {\cal{ O}}(\ep)
{},
\eeq
}
where by $F$ we understand  the {\em exact}  value of  the p-integral $F$
while  the combination  \\  $ \ep^L\,F(\ep, \hat{t}_1,\hat{t}_2, \dots ,\hat{t}_M, \pi)$    should be  
a rational polynomial\footnote{
That is a polynomial having rational coefficients.}    in $\ep,t_1 \dots , t_{M},\pi$.
Every such polynomial is a sum of monomials $T_i$  of the generic form
\beq
\sum_{\alpha} r_{\alpha}T_{\alpha}, \ \  T_{\alpha}=\,\ep^n\, \prod_{i=1,M+1}  t_i^{n_i} 
\EQN{T:def}
{},
\eeq
with $n \le L$, $n_i$ and $r_{\alpha}$   being some non-negative integers and  rational numbers respectively.   
A monomial
$T_\alpha$ will be called  {\em $\pi$-dependent} and denoted as $T_{\pi,\alpha}$ if $n_{M+1} > 0$. Note that a generator
$t_i$ with $i \le M$  may still   include explictly the constant  $\pi$ in its definition, see below.

\ice{ 
while  $\ep^L F(\ep, \hat{t}_1,\hat{t}_2, \dots ,\hat{t}_M, 0,0,  \dots , 0)$ is a polynomial in $t_i$ with coefficients being finite polynomials  in $\ep$ over the field of the rational  numbers.  
}
\
(ii)
For every $t_i$ with $i \le M$  let us define  its hatted counterpart as follows: 
\beq
\hat{t}_i =  t_i + \sum_{j=1,M}  h_j(\ep) \,\,T_{\pi,j}  
\EQN{hj:def}
{}, 
\eeq
with $\{h_j\}$ being rational polynomials in $\ep$ vanishing in the limit of
$\ep=0$ and $T_{\pi,j}$  are {\em all}  $\pi$-dependent monomials as defined in \re{T:def}. 
Then there should exist  a choice of both  a basis ${\cal T}$ and  polynomials $\{h_j\}$ such that for every  
L-loop p-integral  $F(\ep, t_i)$ the following equality holds:
\beq
F(\ep, t_1,t_2,\dots ,  t_{M+1}) = F(\ep, \hat{t}_1,\hat{t}_2, \dots , \hat{t}_M, 0) + {\cal{ O}}(\ep)
\EQN{hat:master}
{}.
\eeq
\ice{
We will call a (renp-function $F$ and/or an AD $\gamma$ {\em $\pi$-free} (both are assumed to b eif they
belong to a ring formed by generators $\{t_i,\ i= 1, \dots , M\}$ over the
field of rational numbers.
}
We will  call  {\em $\pi$-free } 
any polynomial (with possibly $\ep$-dependent coefficients) in  $\{t_i,\ i= 1, \dots , M\}$.   

As  we  will discuss below the sets ${\cal P}_i$ with $i=3,4,5$ are for sure $\pi$-safe while 
${\cal P}_6$  highly likely shares the property.
In what follows we will always assume that
every (renormalized)  L-loop p-function as well as (L+1)-loop $\msbar$ $\beta$-functions and 
anomalous dimensions  are all  expressed in terms of the  generators $t_1,t_2,\dots , t_{M+1}$.

Moreover,  for any   polynomial $P(t_1,t_2,\dots , t_{M+1})$  we define its {\em hatted}
version as
\[
\hat{P}(\hat{t}_1,\hat{t}_2,\dots , \hat{t}_{M}) \defas P(\hat{t}_1,\hat{t}_2,\dots , \hat{t}_{M}, 0) 
{}.
\]
Let 
$F_L$
is a (renormalized, with $\ep$ set to zero) p-function, $\gamma_L$ and $\beta_L$ 
are  the  corresponding anomalous dimension and  the $\beta$-function (all taken in the $L$-loop
approximation).
The following statements have been proved in \cite{Baikov:2018wgs} {\em under the 
condition that the  set ${\cal P}_L$ is $\pi$-safe and that both the  set ${\cal T}$ and 
the polynomilas $\{h_i(\ep)\}$ are fixed}.

\vglue 0.5cm
\noindent
{\bf 1.  No-$\pi$ Theorem } 
\\
(a)
$F_L$
is $p$-free in any (massless) renormalization scheme for which  corresponding   $\beta$-function 
and AD $\g$ are both $\pi$-free at least at the level of $L+1$  loops.  
\\
(b)
The scale-invariant 
combination  $F_{L+1}^{\mathrm si}$ is $\pi$-free in  
any (massless) renormalization scheme provided  the   $\beta$-function
is   $\pi$-independent  at least at the level of $L+1$ loops. 
 
\vglue 0.2cm

\vglue 0.5cm
\noindent
{\bf 2. $\pi$-dependence of L-loop p-functions  }
\\
If $F_L$ is renormalized in $\msbar$-scheme, then all its $\pi$-dependent
contributions can be expressed
in terms of $\hat{F}_{L-1}|_{\ep=0}$,
$\hat{\beta}_{L-1}|_{\ep=0}$ and $\hat{\gamma}_{L-1}|_{\ep=0}$.

\vglue 0.5cm
\noindent
{\bf 3. $\pi$-dependence of L-loop $\beta$-functions and AD  }  
\\
If $\beta_L$ and $\gamma_L$ are given in the $\msbar$-scheme, then all their
$\pi$-dependent contributions can be expressed in terms of
$\hat{\beta}_{L-1}|_{\ep=0}$ and $\hat{\beta}_{L-1}|_{\ep=0}$,
$\hat{\gamma}_{L-1}|_{\ep=0}$ correspondingly.

\section{$\pi$-structure of 3,4,5 and 6-loop p-integrals}

A hatted representation of  p-integrals is  known for  loop numbers $L=3$ \cite{Broadhurst:1999xk},
$L=$4 \cite{Baikov:2010hf}
and 
$L=5$ \cite{Georgoudis:2018olj}. 
In all three cases it was constructed by looking for such a basis ${\cal T}$ as
well as  polynomials $h_j(\ep)$ (see eq. \re{hj:def})  that
eq. \re{hat:master} would be valid for sufficiently  large subset of $ {\cal P}_L$.
\ice{
In all three cases it was constructed by  tuning  polynomials $h_\alpha(\ep)$
(see eq. \re{hj:def}) so that  eq. \re{hat:master} would be valid for all corresponding  master integrals.
}

In principle, \ice{for a loop number $L$}  the strategy requires the
knowledge of all (or almost all) L-loop master integrals.  On the other hand,
if we {\em assume} the $\pi$-safeness of the set ${\cal P}_6$ we could try to
fix polynomials $h_j(\ep)$ by considering some limited subset of ${\cal P}_6$.

Actually, we  do have at  our disposal a subset of ${\cal P}_6$ due to work \cite{Lee:2011jt}
where all 4-loop master integrals  have been computed up to the transcendental  weight  12 in their $\ep$ expansion.
As every particular 4-loop p-integral  divided by $\ep^n$ can be considered as a $(4+n)$ loop p-integral
we have  tried  this subset for  n=$2$. Our results are given  below
(we use even  the  zetas
\ice{
 Zeta[4]                                                                                                                          
          4
        Pi
Out[1]= ---
        90

In[2]:= Zeta[6]//InputForm    
                              
Out[2]//InputForm= Pi^6/945   

In[3]:= Zeta[8]//InputForm    
                              
Out[3]//InputForm= Pi^8/9450  

In[4]:= Zeta[10]//InputForm   
                              
Out[4]//InputForm= Pi^10/93555
}
$\z_4 =\pi^2/90,$ $ \z_6 =  \pi^6/945, \z_8 = \pi^8/9450$ and $\z_{10}=  \pi^{10}/93555$
instead of  the corresponding  even powers of $\pi$).
\begin{flalign}
\unb{\hat{\zeta}_3 \defas \fbox{$\zeta_3$} + \frac{3 \epsilon}{2} \zeta_4}_{L=3} \qquad  \unb{- \frac{5 \epsilon^3}{2} \zeta_6}_{\delta (L=4)}
\qquad  \unb{+\frac{21 \epsilon^5}{2} \zeta_8}_{\delta (L=5)} 
\qquad
\unb{-\frac{153\ep^7}{2} \z_{10}}_{\delta(L=6)}
{},
&&
\EQN{hz3}
\end{flalign}
\vspace{-3mm}
\bfl
\unb{\hat{\zeta}_5  \defas \fbox{$\zeta_5$} + \frac{5 \epsilon}{2} \zeta_6}_{(L=4)} \qquad 
\unb{- \frac{35 \epsilon^3}{4} \zeta_8}_{\delta (L=5)} 
\qquad \unb{ +63 \ep^5 \z_{10} }_{\delta(L=6)}
{}, &&
\EQN{hz5}
\end{flalign}
\vspace{-3mm}
\bfl
\unb{\hat{\zeta}_7   \defas \fbox{$\zeta_7$}}_{L=4} \qquad 
\unb{+ \frac{7 \epsilon}{2} \zeta_8}_{\delta(L=5)}
\qquad \unb{- 21 \ep^3 \z_{10}}_{\delta(L=6)}  
{},&&
\EQN{hz7}
\end{flalign}
\vspace{-3mm}
\bfl 
\unb{
\hat{\varphi} 
\defas \fbox{$\varphi$}
 -3\epsilon\, \zeta_4 \,\zeta_5 + \frac{5 \epsilon}{2} \zeta_3\, \zeta_6 }_{L=5}
\qquad
\unb{
- \frac{24\,\ep^2}{47}\z_{10} + \ep^3\,( - \frac{35}{4}\z_{3}\z_{8} + 5\z_{5}\z_{6})
}_{\delta(L=6)}
{},
 &&
\EQN{hphi}
\end{flalign}
\vspace{-3mm}
\bfl            
\unb{\hat{\z_9} \defas \fbox{$\zeta_9$}}_{L=5} \qquad \unb{+ \frac{9}{2} \ep\,  \z_{10}}_{\delta(L=6)}
{},
&&
\EQN{hz9}
\end{flalign}
\vspace{-3mm}
\bfl
 \unb{\hat{\z}_{7,3}\defas \fbox{$\z_{7,3}  - \frac{793}{94}\z_{10}$}   + 3\ep ( - 7\z_{4}\z_{7} - 5\z_{5}\z_{6})}_{L=6}
{},
 &&
\EQN{hz7z3}
\end{flalign}
\vspace{-3mm}
\bfl
\unb{\hat{\z}_{11}\defas \fbox{$\z_{11}$}
}_{L=6}
{},
 &&
\EQN{hz11}
\end{flalign}
\vspace{-3mm}
\bfl
\unb{\hat{\z}_{5,3,3} \defas \fbox{$\z_{5,3,3}  + 45\z_{2}\z_{9} + 3\z_{4}\z_{7} -\frac{5}{2}\z_{5}\z_{6}$ }}_{L=6}
{}.
 &&
\EQN{hz5z3z3}
\end{flalign}
Here 
\ice{$\varphi=\mzv{6,2}- \mzv{3,5} \approx -0.1868414$ }
\beq
\varphi \defas \frac{3}{5}\, \z_{5,3} + \z_3\, \z_5 -\frac{29}{20} \,\z_8 
= \mzv{6,2}- \mzv{3,5}
\approx -0.1868414
\EQN{varphi}
\eeq
and   
multiple zeta values are defined  as \cite{Blumlein:2009cf}
\beq
\mzv{n_1,n_2} \defas \sum_{i > j > 0 } \frac{1}{i^{n_1} j^{n_2}}, \ \
\mzv{n_1,n_2,n_3} \defas \sum_{i > j >  k > 0 } \frac{1}{i^{n_1} j^{n_2} k^{n_3}  }
\EQN{mzv}
{}.
\eeq
Some comments on these eqs. are  in order.

\begin{itemize}

\item The boxed entries form a set of $\pi$-independent (by definition!) generators  for
the cases of $L=3$ (eq. \re{hz3}), $L=4$ (eqs. (\ref{hz3}---\ref{hz7}),  
$L=5$ (eqs. (\ref{hz3}--\ref{hz9}) and  $L=6$
(eqs. (\ref{hz3}---\ref{hz5z3z3}).

\item For $L=5$ we recover  the
hatted representation  
for the set ${\cal P}_5$ first found  in \cite{Georgoudis:2018olj}.

\item  We do not claim that the generators  
\beq
\z_3,\z_5,\z_7, \phi, \z_9, \hz_{7,3}|_{\ep=0},
\hz_{5,3,3} \ \mbox{and} \ \ \pi
\EQN{gens}
\eeq
are sufficient to  present the pole and finite parts of every
6-loop p-integral. In fact, it is not true 
\cite{Broadhurst:1995km,Panzer:2015ida,Schnetz:2016fhy}. 
However we believe \ice{think} that it is safe to assume
that all missing irrational constants  can be associated with the values of some 
convergent 6-loop p-integrals at $\ep=0$. 

\end{itemize}

\section{$\pi$-dependence of 7-loop $\beta$-functions and AD}

Using the approach of \cite{Baikov:2018wgs} and the hatted representation of
the irrational generators \re{gens} as described by eqs. \re{hz3}-\re{hz5z3z3}
\ice{for the 6-loop case} we can straightforwardly predict the $\pi$-dependent
terms in the $\beta$-function and the anomalous dimensions in the case of {\em
  any} 1-charge minimally renormalized field model at the level of 7 loops.

Our results read (the
combination $F^{ t_{\alpha_1} t_{\alpha_2}\dots  t_{\alpha_n}}$ stands for
the coefficient of the monomial \\ $(t_{\alpha_1} t_{\alpha_2}\dots  t_{\alpha_n})$ in $F$; in addition,  by
$F^{(1)}$ we understand $F$ with every  generator $t_i$  from $\{t_1, t_2,  \dots , t_{M+1}\}$ set to zero).
 \begin{flalign} 
  \gamma_4^{\z_4} =  
       -\frac{1}{2}\,  \beta_3^{\z_3}  \gamma_1    
       +\frac{3}{2}\,  \beta_1  \gamma_3^{\z_3}    
      ,
 &&\EQN{g41} \end{flalign} \vspace{\myl}

 \begin{flalign} 
  \gamma_5^{\z_4} =  
       -\frac{3}{8}\,  \beta_4^{\z_3}  \gamma_1    
       +\frac{3}{2}\,  \beta_2  \gamma_3^{\z_3}    
       - \beta_3^{\z_3}  \gamma_2    
       +\frac{3}{2}\,  \beta_1  \gamma_4^{\z_3}    
      ,
 &&\EQN{g51} \end{flalign} \vspace{\myl}

 \begin{flalign} 
  \gamma_5^{\z_6} =  
       -\frac{5}{8}\,  \beta_4^{\z_5}  \gamma_1    
       +\frac{5}{2}\,  \beta_1  \gamma_4^{\z_5}    
      ,
 &&\EQN{g52} \end{flalign} \vspace{\myl}

 \begin{flalign} 
  \gamma_5^{\z_3 \z_4} = 0,
 &&\EQN{g53} \end{flalign} \vspace{\myl}

 \begin{flalign} 
  \gamma_6^{\z_4}   =  \frac{3}{2}\,  \beta^{(1)}_3  \gamma_3^{\z_3}    
       -\frac{3}{10}\,  \beta_5^{\z_3}  \gamma_1    
       -\frac{3}{4}\,  \beta_4^{\z_3}  \gamma_2    
       +\frac{3}{2}\,  \beta_2  \gamma_4^{\z_3}    
       -\frac{3}{2}\,  \beta_3^{\z_3}  \gamma^{(1)}_3    
       +\frac{3}{2}\,  \beta_1  \gamma_5^{\z_3}    
      ,
 &&\EQN{g61} \end{flalign} \vspace{\myl}

 \begin{flalign} 
  \gamma_6^{\z_6} =  
       -\frac{1}{2}\,  \beta_5^{\z_5}  \gamma_1    
       -\frac{5}{4}\,  \beta_4^{\z_5}  \gamma_2    
       +\frac{5}{2}\,  \beta_2  \gamma_4^{\z_5}    
       +\frac{5}{2}\,  \beta_1  \gamma_5^{\z_5}    
       +\frac{3}{2}\,  \beta_1^2  \beta_3^{\z_3}  \gamma_1    
       -\frac{5}{2}\,  \beta_1^3  \gamma_3^{\z_3}    
      ,
 &&\EQN{g62} \end{flalign} \vspace{\myl}

 \begin{flalign} 
  \gamma_6^{\z_3 \z_4} =  
       -\frac{3}{5}\,  \beta_5^{\z_3^2}  \gamma_1    
       +3  \beta_1  \gamma_5^{\z_3^2}    
      ,
 &&\EQN{g63} \end{flalign} \vspace{\myl}

 \begin{flalign} 
  \gamma_6^{\z_8} =  
       -\frac{7}{10}\,  \beta_5^{\z_7}  \gamma_1    
       +\frac{7}{2}\,  \beta_1  \gamma_5^{\z_7}    
      ,
 &&\EQN{g64} \end{flalign} \vspace{\myl}

 \begin{flalign} 
  \gamma_6^{\z_3 \z_6} = \gamma_6^{\z_4 \z_5} = 0,
 &&\EQN{g65} \end{flalign} \vspace{\myl} 
  
 \ice{ 
 \begin{flalign} 
  \gamma_6^{\z_4 \z_5} = 0,
 &&\EQN{} \end{flalign} \vspace{\myl} 
  }

 \begin{flalign} 
 \gamma_7^{\z_4} =  
       &-\frac{1}{4}\,  \beta_6^{\z_3}  \gamma_1    
       +\frac{3}{2}\,  \beta^{(1)}_3  \gamma_4^{\z_3}    
       +\frac{3}{2}\,  \beta^{(1)}_4  \gamma_3^{\z_3}  
      -\frac{3}{5}\,  \beta_5^{\z_3}  \gamma_2    \hspace{73mm}
\nnb
\\
      &    -\frac{9}{8}\,  \beta_4^{\z_3}  \gamma^{(1)}_3    
       +\frac{3}{2}\,  \beta_2  \gamma_5^{\z_3}    
       -2  \beta_3^{\z_3}  \gamma^{(1)}_4    
       +\frac{3}{2}\,  \beta_1  \gamma_6^{\z_3}    
      ,
\EQN{g71}  
\end{flalign}
\vspace{\myl}

 \begin{flalign} 
  \gamma_7^{\z_6} = &  
       -\frac{5}{12}\,  \beta_6^{\z_5}  \gamma_1    
       +\frac{5}{2}\,  \beta^{(1)}_3  \gamma_4^{\z_5}    
       - \beta_5^{\z_5}  \gamma_2    
       -\frac{15}{8}\,  \beta_4^{\z_5}  \gamma^{(1)}_3    
       +\frac{5}{2}\,  \beta_2  \gamma_5^{\z_5}    +\frac{5}{2}\,  \beta_1  \gamma_6^{\z_5}    
 \hspace{73mm}
\nnb
 \\
 &+\frac{5}{2}\,  \beta_1  \beta_3^{\z_3}  \beta_2  \gamma_1    
       +\frac{5}{4}\,  \beta_1^2  \beta_4^{\z_3}  \gamma_1    
       -\frac{15}{2}\,  \beta_1^2  \beta_2  \gamma_3^{\z_3}    
       +3  \beta_1^2  \beta_3^{\z_3}  \gamma_2    
       -\frac{5}{2}\,  \beta_1^3  \gamma_4^{\z_3}    
     ,
\EQN{g72} \end{flalign} \vspace{\myl}

 \begin{flalign} 
  \gamma_7^{\z_3 \z_4} =  
       -\frac{1}{2}\,  \beta_6^{\z_3^2}  \gamma_1    
       -\frac{6}{5}\,  \beta_5^{\z_3^2}  \gamma_2    
       +\frac{3}{8}\,  \beta_4^{\z_3}  \gamma_3^{\z_3}    
       +3  \beta_2  \gamma_5^{\z_3^2}    
       -\frac{1}{2}\,  \beta_3^{\z_3}  \gamma_4^{\z_3}    
       +3  \beta_1  \gamma_6^{\z_3^2}    
      ,
 &&\EQN{g73} \end{flalign} \vspace{\myl}

 \begin{flalign} 
  \gamma_7^{\z_8} =&  
       -\frac{7}{12}\,  \beta_6^{\z_7}  \gamma_1    
       -\frac{7}{5}\,  \beta_5^{\z_7}  \gamma_2    
       +\frac{7}{2}\,  \beta_2  \gamma_5^{\z_7}    
       +\frac{7}{12}\, (\beta_3^{\z_3})^2  \gamma_1    
       +\frac{7}{2}\,  \beta_1  \gamma_6^{\z_7}    
      -\frac{7}{8}\,  \beta_1  \beta_5^{\z_3^2}  \gamma_1   
\nnb
\\
       &-\frac{7}{8}\,  \beta_1  \beta_3^{\z_3}  \gamma_3^{\z_3}    
       +\frac{21}{8}\,  \beta_1^2  \gamma_5^{\z_3^2}    
       +\frac{35}{8}\,  \beta_1^2  \beta_4^{\z_5}  \gamma_1    
       -\frac{35}{4}\,  \beta_1^3  \gamma_4^{\z_5}    
      ,
&&\EQN{g74} \end{flalign} \vspace{\myl}

 \begin{flalign} 
  \gamma_7^{\z_3 \z_6} =  
       -\frac{5}{12}\,  \beta_6^{\z_3 \z_5}  \gamma_1    
       -\frac{5}{12}\, \beta_6^{\phi}   \gamma_1    
       -\frac{15}{8}\,  \beta_4^{\z_5}  \gamma_3^{\z_3}    
       +\frac{5}{2}\,  \beta_3^{\z_3}  \gamma_4^{\z_5}    
       +\frac{5}{2}\,  \beta_1  \gamma_6^{\z_3 \z_5}    
       +\frac{5}{2}\,  \beta_1 \gamma_6^{\phi}     
      ,
 &&\EQN{g75} \end{flalign} \vspace{\myl}

 \begin{flalign} 
  \gamma_7^{\z_4 \z_5} =  
       -\frac{1}{4}\,  \beta_6^{\z_3 \z_5}  \gamma_1    
       +\frac{1}{2}\, \beta_6^{\phi}   \gamma_1    
       +\frac{3}{2}\,  \beta_4^{\z_5}  \gamma_3^{\z_3}    
       -2  \beta_3^{\z_3}  \gamma_4^{\z_5}    
       +\frac{3}{2}\,  \beta_1  \gamma_6^{\z_3 \z_5}    
       -3  \beta_1 \gamma_6^{\phi}     
      ,
 &&\EQN{g76} \end{flalign} \vspace{\myl}

 \begin{flalign} 
  \gamma_7^{\z_{10}} =  
       -\frac{3}{4}\,  \beta_6^{\z_9}  \gamma_1    
       +\frac{9}{2}\,  \beta_1  \gamma_6^{\z_9}    
      ,
 &&\EQN{g77} \end{flalign} \vspace{\myl}

 \begin{flalign} 
  \gamma_7^{\z_4 \z_3^2} =  
       -\frac{3}{4}\,  \beta_6^{\z_3^3}  \gamma_1    
       +\frac{9}{2}\,  \beta_1  \gamma_6^{\z_3^3}    
      ,
 &&\EQN{g78} \end{flalign} \vspace{\myl}

 \begin{flalign} 
  \gamma_7^{\z_4 \z_7} =\gamma_7^{\z_5 \z_6}=\gamma_7^{\z_3 \z_8} =  0
{}.
 &&\EQN{g79} \end{flalign}

\ice{
 \begin{flalign} 
  \gamma_7^{\z_4 \z_7} = 0,
 &&\EQN{g7} \end{flalign} \vspace{\myl}

 \begin{flalign} 
  \gamma_7^{\z_5 \z_6} = 0,
 &&\EQN{g7} \end{flalign} \vspace{\myl}

 \begin{flalign} 
  \gamma_7^{\z_3 \z_8} = 0,
 &&\EQN{g7} \end{flalign} \vspace{\myl} 
}

The results for $\pi$-dependent contributions to a  $\beta$-function are obtained from
the above eqs. by a formal replacement of $\gamma$ by $\beta$ in  every term. For instance, the 
7-loop $\pi$-dependent contributions read: 
 \begin{flalign} 
  \beta_7^{\z_4}   =  \frac{3}{8}\,  \beta_4^{\z_3}  \beta^{(1)}_3    
       +\frac{9}{10}\,  \beta_2  \beta_5^{\z_3}    
       -\frac{1}{2}\,  \beta_3^{\z_3}  \beta^{(1)}_4    
       +\frac{5}{4}\,  \beta_1  \beta_6^{\z_3}    
      ,
 &&\EQN{b71} \end{flalign} \vspace{\myl}

 \begin{flalign} 
  \beta_7^{\z_6}   =  \frac{5}{8}\,  \beta_4^{\z_5}  \beta^{(1)}_3    
       +\frac{3}{2}\,  \beta_2  \beta_5^{\z_5}    
       +\frac{25}{12}\,  \beta_1  \beta_6^{\z_5}    
       -2  \beta_1^2  \beta_3^{\z_3}  \beta_2    
       -\frac{5}{4}\,  \beta_1^3  \beta_4^{\z_3}    
      ,
 &&\EQN{b72} \end{flalign} \vspace{\myl} 
  

 \begin{flalign} 
  \beta_7^{\z_3 \z_4}   =  \frac{9}{5}\,  \beta_2  \beta_5^{\z_3^2}    
       -\frac{1}{8}\,  \beta_3^{\z_3}  \beta_4^{\z_3}    
       +\frac{5}{2}\,  \beta_1  \beta_6^{\z_3^2}    
      ,
 &&\EQN{b73} \end{flalign} \vspace{\myl} 
  

 \begin{flalign} 
  \beta_7^{\z_8}   =  \frac{21}{10}\,  \beta_2  \beta_5^{\z_7}    
       +\frac{35}{12}\,  \beta_1  \beta_6^{\z_7}    
       -\frac{7}{24}\,  \beta_1 (\beta_3^{\z_3})^2    
       +\frac{7}{4}\,  \beta_1^2  \beta_5^{\z_3^2}    
       -\frac{35}{8}\,  \beta_1^3  \beta_4^{\z_5}    
      ,
 &&\EQN{b74} \end{flalign} \vspace{\myl} 
  
 \begin{flalign} 
  \beta_7^{\z_3 \z_6}   =  \frac{5}{8}\,  \beta_3^{\z_3}  \beta_4^{\z_5}    
       +\frac{25}{12}\,  \beta_1  \beta_6^{\z_3 \z_5}    
       +\frac{25}{12}\,  \beta_1 \beta_6^{\phi}     
      ,
 &&\EQN{b75} \end{flalign} \vspace{\myl}

 \begin{flalign} 
  \beta_7^{\z_4 \z_5} =  
       -\frac{1}{2}\,  \beta_3^{\z_3}  \beta_4^{\z_5}    
       +\frac{5}{4}\,  \beta_1  \beta_6^{\z_3 \z_5}    
       -\frac{5}{2}\,  \beta_1 \beta_6^{\phi}     
      ,
 &&\EQN{b76} \end{flalign} \vspace{\myl}

 \begin{flalign} 
  \beta_7^{\z_{10}}   =  \frac{15}{4}\,  \beta_1  \beta_6^{\z_9}    
      ,
 &&\EQN{b77} \end{flalign} \vspace{\myl}

 \begin{flalign} 
  \beta_7^{\z_4 \z_3^2}   =  \frac{15}{4}\,  \beta_1  \beta_6^{\z_3^3}    
      ,
 &&\EQN{b78} \end{flalign} \vspace{\myl} 
  
  \begin{flalign} 
  \beta_7^{\z_4 \z_7} = \beta_7^{\z_5 \z_6}=  \beta_7^{\z_3 \z_8} = 0.
 &&\EQN{b79} \end{flalign}

\subsection{Tests} 

With eqs. \re{g41}--\re{b79} we have been able to reproduce  successfully  all $\pi$-dependent 
constants appearing in the $\beta$-function and anomalous dimensions $\gamma_m$ and $\gamma_2$ of the
$O(n)$ $\varphi^4$ model which all are known at 7 loops from \cite{Schnetz:2016fhy}. In
addition, we have checked that the $\pi$-dependent contributions to the  terms of
order $n_f^6\alpha_s^7$ in the the QCD $\beta$-function as well as to the terms of order
$n_f^6\alpha_s^7$ and of order $n_f^5\alpha_s^7$ contributing to the quark mass AD 
(all computed in \cite{Gracey:1996he,Ciuchini:1999cv,Ciuchini:1999wy}) are in agreement with
constraints \ice{relations} \re{b71}--\re{b79} and \re{g71}--\re{g79} respectively.

Numerous successful tests  at  4,5 and 6 loops 
have been presented in \cite{Baikov:2018wgs}.


\acknowledgments 
We are grateful to E. Panzer and V. Smirnov for useful discussions and  good advice.

The work of P.A.~Baikov is supported in part by the 
grant RFBR 17-02-00175A of the Russian Foundation for Basic Research.
The work by K. G. Chetykin was supported by 
the German Federal Ministry for Education and Research BMBF
through Grant  No. \ice{05H2015  and}05H15GUCC1.

\providecommand{\href}[2]{#2}\begingroup\raggedright\endgroup

\ed

\mbib

\ed

\ed
\ice{
 \begin{flalign} 
  \beta_4^{\z_4}   =   \beta_1  \beta_3^{\z_3}    
      ,
 &&\EQN{g7} \end{flalign} \vspace{\myl} 
 \begin{flalign} 
  \beta_5^{\z_4}   =  \frac{1}{2}\,  \beta_3^{\z_3}  \beta_2    
       +\frac{9}{8}\,  \beta_1  \beta_4^{\z_3}    
      ,
 &&\EQN{g7} \end{flalign} \vspace{\myl}

 \begin{flalign} 
  \beta_5^{\z_6}   =  \frac{15}{8}\,  \beta_1  \beta_4^{\z_5}    
      ,
 &&\EQN{g7} \end{flalign} \vspace{\myl}

 \begin{flalign} 
  \beta_5^{\z_3 \z_4} = 0,
 &&\EQN{g7} \end{flalign} \vspace{\myl}

 \begin{flalign} 
  \beta_6^{\z_4}   =  \frac{3}{4}\,  \beta_2  \beta_4^{\z_3}    
       +\frac{6}{5}\,  \beta_1  \beta_5^{\z_3}    
      ,
 &&\EQN{g7} \end{flalign} \vspace{\myl}

 \begin{flalign} 
  \beta_6^{\z_6}   =  \frac{5}{4}\,  \beta_2  \beta_4^{\z_5}    
       +2  \beta_1  \beta_5^{\z_5}    
       - \beta_1^3  \beta_3^{\z_3}    
      ,
 &&\EQN{g7} \end{flalign} \vspace{\myl}

 \begin{flalign} 
  \beta_6^{\z_3 \z_4}   =  \frac{12}{5}\,  \beta_1  \beta_5^{\z_3^2}    
      ,
 &&\EQN{g7} \end{flalign} \vspace{\myl}

 \begin{flalign} 
  \beta_6^{\z_8}   =  \frac{14}{5}\,  \beta_1  \beta_5^{\z_7}    
      ,
 &&\EQN{g7} \end{flalign} \vspace{\myl}

 \begin{flalign} 
  \beta_6^{\z_3 \z_6} = 0,
 &&\EQN{g7} \end{flalign} \vspace{\myl}

 \begin{flalign} 
  \beta_6^{\z_4 \z_5} = 0,
 &&\EQN{g7} \end{flalign} \vspace{\myl}

 \begin{flalign} 
  \beta_7^{\z_4}   =  \frac{3}{8}\,  \beta_4^{\z_3}  \beta^{(1)}_3    
       +\frac{9}{10}\,  \beta_2  \beta_5^{\z_3}    
       -\frac{1}{2}\,  \beta_3^{\z_3}  \beta^{(1)}_4    
       +\frac{5}{4}\,  \beta_1  \beta_6^{\z_3}    
      ,
 &&\EQN{g7} \end{flalign} \vspace{\myl}

 \begin{flalign} 
  \beta_7^{\z_6}   =  \frac{5}{8}\,  \beta_4^{\z_5}  \beta^{(1)}_3    
       +\frac{3}{2}\,  \beta_2  \beta_5^{\z_5}    
       +\frac{25}{12}\,  \beta_1  \beta_6^{\z_5}    
       -2  \beta_1^2  \beta_3^{\z_3}  \beta_2    
       -\frac{5}{4}\,  \beta_1^3  \beta_4^{\z_3}    
      ,
 &&\EQN{g7} \end{flalign} \vspace{\myl}

 \begin{flalign} 
  \beta_7^{\z_3 \z_4}   =  \frac{9}{5}\,  \beta_2  \beta_5^{\z_3^2}    
       -\frac{1}{8}\,  \beta_3^{\z_3}  \beta_4^{\z_3}    
       +\frac{5}{2}\,  \beta_1  \beta_6^{\z_3^2}    
      ,
 &&\EQN{g7} \end{flalign} \vspace{\myl}

 \begin{flalign} 
  \beta_7^{\z_8}   =  \frac{21}{10}\,  \beta_2  \beta_5^{\z_7}    
       +\frac{35}{12}\,  \beta_1  \beta_6^{\z_7}    
       -\frac{7}{24}\,  \beta_1 (\beta_3^{\z_3})^2    
       +\frac{7}{4}\,  \beta_1^2  \beta_5^{\z_3^2}    
       -\frac{35}{8}\,  \beta_1^3  \beta_4^{\z_5}    
      ,
 &&\EQN{g7} \end{flalign} \vspace{\myl}

 \begin{flalign} 
  \beta_7^{\z_3 \z_6}   =  \frac{5}{8}\,  \beta_3^{\z_3}  \beta_4^{\z_5}    
       +\frac{25}{12}\,  \beta_1  \beta_6^{\z_3 \z_5}    
       +\frac{25}{12}\,  \beta_1 \beta_6^{\phi}     
      ,
 &&\EQN{g7} \end{flalign} \vspace{\myl}

 \begin{flalign} 
  \beta_7^{\z_4 \z_5} =  
       -\frac{1}{2}\,  \beta_3^{\z_3}  \beta_4^{\z_5}    
       +\frac{5}{4}\,  \beta_1  \beta_6^{\z_3 \z_5}    
       -\frac{5}{2}\,  \beta_1 \beta_6^{\phi}     
      ,
 &&\EQN{g7} \end{flalign} \vspace{\myl}

 \begin{flalign} 
  \beta_7^{\z_{10}}   =  \frac{15}{4}\,  \beta_1  \beta_6^{\z_9}    
      ,
 &&\EQN{g7} \end{flalign} \vspace{\myl}

 \begin{flalign} 
  \beta_7^{\z_4 \z_3^2}   =  \frac{15}{4}\,  \beta_1  \beta_6^{\z_3^3}    
      ,
 &&\EQN{g7} \end{flalign} \vspace{\myl} 
  
  \begin{flalign} 
  \beta_7^{\z_4 \z_7} = \beta_7^{\z_5 \z_6} \beta_7^{\z_3 \z_8} = 0,
 &&\EQN{g7} \end{flalign} \vspace{\myl}

\ice{
 \begin{flalign} 
  \beta_7^{\z_4 \z_7} = 0,
 &&\EQN{g7} \end{flalign} \vspace{\myl}

 \begin{flalign} 
  \beta_7^{\z_5 \z_6} = 0,
 &&\EQN{g7} \end{flalign} \vspace{\myl}

 \begin{flalign} 
  \beta_7^{\z_3 \z_8} = 0,
 &&\EQN{g7} \end{flalign} \vspace{\myl} 
  }

}
\begin{thebibliography}{10}

\bibitem{Baikov:2018wgs}
P.~A. Baikov and K.~G. Chetyrkin, \emph{{The structure of generic anomalous
  dimensions and no-$\pi$ theorem for massless propagators}},
  \href{http://dx.doi.org/10.1007/JHEP06(2018)141}{\emph{JHEP} {\bfseries 06}
  (2018) 141}, [\href{https://arxiv.org/abs/1804.10088}{{\ttfamily
  1804.10088}}].

\bibitem{Gorishnii:1991vf}
S.~G. Gorishny, A.~L. Kataev and S.~A. Larin, \emph{The { ${\cal
  O}(\alpha_s^3)$} corrections to {$\sigma_{\rm tot}(e^+ e^- \to {\rm
  hadrons})$} and {$\sigma( {\tau} \to \nu_{\tau} + {\rm hadrons})$} in {
  QCD}}, {\emph{Phys. Lett.} {\bfseries B259} (1991) 144--150}.

\bibitem{Jamin:2017mul}
M.~Jamin and R.~Miravitllas, \emph{{Absence of even-integer $\zeta$-function
  values in Euclidean physical quantities in QCD}},
  \href{http://dx.doi.org/10.1016/j.physletb.2018.02.030}{\emph{Phys. Lett.}
  {\bfseries B779} (2018) 452--455},
  [\href{https://arxiv.org/abs/1711.00787}{{\ttfamily 1711.00787}}].

\bibitem{Davies:2017hyl}
J.~Davies and A.~Vogt, \emph{{Absence of $\pi^2$ terms in physical anomalous
  dimensions in DIS: Verification and resulting predictions}},
  \href{http://dx.doi.org/10.1016/j.physletb.2017.11.036}{\emph{Phys. Lett.}
  {\bfseries B776} (2018) 189--194},
  [\href{https://arxiv.org/abs/1711.05267}{{\ttfamily 1711.05267}}].

\bibitem{Chetyrkin:2017bjc}
K.~G. Chetyrkin, G.~Falcioni, F.~Herzog and J.~A.~M. Vermaseren,
  \emph{{Five-loop renormalisation of QCD in covariant gauges}},
  \href{http://dx.doi.org/10.1007/JHEP12(2017)006,
  10.1007/JHEP10(2017)179}{\emph{JHEP} {\bfseries 10} (2017) 179},
  [\href{https://arxiv.org/abs/1709.08541}{{\ttfamily 1709.08541}}].

\bibitem{Ruijl:2018poj}
B.~Ruijl, F.~Herzog, T.~Ueda, J.~A.~M. Vermaseren and A.~Vogt,
  \emph{{R*-operation and five-loop calculations}},
  \href{http://dx.doi.org/10.22323/1.290.0011}{\emph{PoS} {\bfseries
  RADCOR2017} (2018) 011}, [\href{https://arxiv.org/abs/1801.06084}{{\ttfamily
  1801.06084}}].

\bibitem{Baikov:2016tgj}
P.~A. Baikov, K.~G. Chetyrkin and J.~H. K{\"u}hn, \emph{{Five-Loop Running of
  the QCD coupling constant}},
  \href{http://dx.doi.org/10.1103/PhysRevLett.118.082002}{\emph{Phys. Rev.
  Lett.} {\bfseries 118} (2017) 082002},
  [\href{https://arxiv.org/abs/1606.08659}{{\ttfamily 1606.08659}}].

\bibitem{Herzog:2017ohr}
F.~Herzog, B.~Ruijl, T.~Ueda, J.~A.~M. Vermaseren and A.~Vogt, \emph{{The
  five-loop beta function of Yang-Mills theory with fermions}},
  \href{http://dx.doi.org/10.1007/JHEP02(2017)090}{\emph{JHEP} {\bfseries 02}
  (2017) 090}, [\href{https://arxiv.org/abs/1701.01404}{{\ttfamily
  1701.01404}}].

\bibitem{Luthe:2017ttg}
T.~Luthe, A.~Maier, P.~Marquard and Y.~Schr{\"o}der, \emph{{The five-loop Beta
  function for a general gauge group and anomalous dimensions beyond Feynman
  gauge}}, \href{http://dx.doi.org/10.1007/JHEP10(2017)166}{\emph{JHEP}
  {\bfseries 10} (2017) 166},
  [\href{https://arxiv.org/abs/1709.07718}{{\ttfamily 1709.07718}}].

\bibitem{Broadhurst:1999xk}
D.~J. Broadhurst, \emph{{Dimensionally continued multiloop gauge theory}},
  \href{https://arxiv.org/abs/hep-th/9909185}{{\ttfamily hep-th/9909185}}.

\bibitem{Baikov:2010hf}
P.~A. Baikov and K.~G. Chetyrkin, \emph{{Four-Loop Massless Propagators: an
  Algebraic Evaluation of All Master Integrals}},
  \href{http://dx.doi.org/10.1016/j.nuclphysb.2010.05.004}{\emph{Nucl. Phys.}
  {\bfseries B837} (2010) 186--220},
  [\href{https://arxiv.org/abs/1004.1153}{{\ttfamily 1004.1153}}].

\bibitem{Georgoudis:2018olj}
A.~Georgoudis, V.~Goncalves, E.~Panzer and R.~Pereira, \emph{{Five-loop
  massless propagator integrals}},
  \href{https://arxiv.org/abs/1802.00803}{{\ttfamily 1802.00803}}.

\bibitem{Lee:2011jt}
R.~N. Lee, A.~V. Smirnov and V.~A. Smirnov, \emph{{Master Integrals for
  Four-Loop Massless Propagators up to Transcendentality Weight Twelve}},
  \href{http://dx.doi.org/10.1016/j.nuclphysb.2011.11.005}{\emph{Nucl. Phys.}
  {\bfseries B856} (2012) 95--110},
  [\href{https://arxiv.org/abs/1108.0732}{{\ttfamily 1108.0732}}].

\bibitem{Blumlein:2009cf}
J.~Blumlein, D.~J. Broadhurst and J.~A.~M. Vermaseren, \emph{{The Multiple Zeta
  Value Data Mine}},
  \href{http://dx.doi.org/10.1016/j.cpc.2009.11.007}{\emph{Comput. Phys.
  Commun.} {\bfseries 181} (2010) 582--625},
  [\href{https://arxiv.org/abs/0907.2557}{{\ttfamily 0907.2557}}].

\bibitem{Broadhurst:1995km}
D.~J. Broadhurst and D.~Kreimer, \emph{Knots and numbers in phi**4 theory to 7
  loops and beyond}, {\emph{Int. J. Mod. Phys.} {\bfseries C6} (1995)
  519--524}, [\href{https://arxiv.org/abs/hep-ph/9504352}{{\ttfamily
  hep-ph/9504352}}].

\bibitem{Panzer:2015ida}
E.~Panzer, \emph{{Feynman integrals and hyperlogarithms}}.
\newblock PhD thesis, Humboldt U., Berlin, Inst. Math., 2015.
\newblock \href{https://arxiv.org/abs/1506.07243}{{\ttfamily 1506.07243}}.
\newblock 10.18452/17157.

\bibitem{Schnetz:2016fhy}
O.~Schnetz, \emph{{Numbers and Functions in Quantum Field Theory}},
  \href{http://dx.doi.org/10.1103/PhysRevD.97.085018}{\emph{Phys. Rev.}
  {\bfseries D97} (2018) 085018},
  [\href{https://arxiv.org/abs/1606.08598}{{\ttfamily 1606.08598}}].

\bibitem{Gracey:1996he}
J.~Gracey, \emph{{The QCD Beta function at ${\cal O}(1/N_f)$}},
  \href{http://dx.doi.org/10.1016/0370-2693(96)00105-0}{\emph{Phys.Lett.}
  {\bfseries B373} (1996) 178--184},
  [\href{https://arxiv.org/abs/hep-ph/9602214}{{\ttfamily hep-ph/9602214}}].

\bibitem{Ciuchini:1999cv}
M.~Ciuchini, S.~E. Derkachov, J.~Gracey and A.~Manashov, \emph{{Quark mass
  anomalous dimension at O(1/N(f)**2) in QCD}},
  \href{http://dx.doi.org/10.1016/S0370-2693(99)00573-0}{\emph{Phys.Lett.}
  {\bfseries B458} (1999) 117--126},
  [\href{https://arxiv.org/abs/hep-ph/9903410}{{\ttfamily hep-ph/9903410}}].

\bibitem{Ciuchini:1999wy}
M.~Ciuchini, S.~E. Derkachov, J.~Gracey and A.~Manashov, \emph{{Computation of
  quark mass anomalous dimension at O(1 / N**2(f)) in quantum chromodynamics}},
  \href{http://dx.doi.org/10.1016/S0550-3213(00)00209-1}{\emph{Nucl.Phys.}
  {\bfseries B579} (2000) 56--100},
  [\href{https://arxiv.org/abs/hep-ph/9912221}{{\ttfamily hep-ph/9912221}}].

\end{thebibliography}
